\documentclass[aps,prd,onecolumn,groupedaddress,showpacs,nofootinbib,amssymb]{revtex4-2}
\usepackage[dvips]{graphicx}
\usepackage{amssymb}
\usepackage{amsmath}
\usepackage{graphicx}
\usepackage{amsfonts}
\usepackage{bm}
\usepackage{latexsym,float}
\usepackage{epsfig}
\usepackage{epstopdf}
\usepackage{amssymb}

\usepackage{verbatim}
\usepackage{multirow}
\usepackage{color}
\newcommand{\mycomment}[1]{}
\begin{document}

\title{On Isaacson's Procedure in $f(R)$-Gravity with Second Order Metric Perturbations}

\author{Petr V. Tretyakov}
\email{tpv@theor.jinr.ru} \affiliation{Joint Institute for
Nuclear Research, Joliot-Curie 6, 141980 Dubna, Moscow region,
Russia}

\author{Alexander N. Petrov}
\email{alex.petrov55@gmail.com} \affiliation{Sternberg Astronomical Institute, Moscow M.V. Lomonosov State University, \\ Universitetskii pr. 13, 119234 Moscow,
Russia}

\begin{abstract}
In the recent paper \cite{TP1} we generalized classical Isaacson's procedure for describing back-reaction of the averaged energy-momentum for high frequency gravitational waves to $f(R)$ gravity case with a choice of the de Sitter background. To describe this second order effect we used standard first order decomposition of metric tensor. In this paper, we take into account possible contribution to this process from second order term in metric tensor decomposition. We find that on the de Sitter background all possible contributions from this term are strongly suppressed and therefore results obtained in the paper \cite{TP1} are left unchanged.
\end{abstract}


\maketitle

\section{Introduction}

Classical Isaacson's approach \cite{A1,A2} to construct the averaged energy-momentum tensor for high frequency gravitational waves is well known and described in canonical textbooks on general relativity (GR), see, for example, \cite{LL}, and in more detail in \cite{MTW}. It implies that in high frequency limit small ripple produces non-trivial back-reaction on the background metric under the Brill-Hartle \cite{BH1} averaging procedure and this process may present some effective energy-momentum tensor for gravitational radiation. In the light of rising popularity of modified gravity theories  there was a number of attempts to generalize this approach to such theories. More concretely, see \cite{SY1,PM1,P1,LC1,DLLL1} and references therein.

The key point in Isaacson's procedure is the analysis of the scales. Due to specific choice of scales it's occur that the leading terms in decomposition of vacuum Einstein equation are the first order terms, which have the magnitude of $\epsilon^{-1}$ with $\epsilon \ll 1$. Physically it corresponds to the wave equation. The background (zero) and second terms have the similar order $O(1)$, and exactly this fact allows us to interpret the second order terms after the averaging as the source that bends  a spacetime.

In our previous paper \cite{TP1}, we develop Isaacson's mechanism in context of $f(R)$-gravity.  As an initial background, we consider the de Sitter (dS) space. Thus, at least, we introduce an additional scale --- the dS radius. Besides, $f(R)$-gravity, including higher derivative terms, is more complicated theory with respect to the usual GR. By these arguments, we encountered with some criticism in partial order that further terms in decomposition of metric may have influence on the studying process. Indeed for the outside observer, which is not deeply immersed in the subject, all of these looks as a reasonable one. By this reason, we decided to present this short report, which sheds light on this question and finalized the study in \cite{TP1}.

The paper is organized as follows. In Sec.\ref{sect.2} we discuss the scales in classical Isaacson's approach, the difference with our approach and introduce second order of metric perturbations.  In Sec.\ref{sect.3} we present second order term if decomposition of modified Einstein equation, calculate carefully all contributions from the second order of metric perturbations and analyse its influence on energy-momentum tensor. In Sec.\ref{sect.4} we present some concluding remarks.

\section{Preliminaries }
\label{sect.2}

\subsection{Classical Isaacson's decomposition of metric, and scales in high-frequency limit}

In the original study \cite{A1,A2} Isaacson  has suggested a scheme of high frequency limit. Gravitational waves are described by perturbations in
\begin{equation}
g_{ik}=g_{ik}^{B}+ h_{ik},
\label{decomposition}
\end{equation}
where small Latin indexes numerate spacetime coordinates $i,k,\ldots = 0,1,2,3$. It is assumed that $g^B_{ik} = O(1)$, whereas the amplitude of $h_{ik} = O(\epsilon)$  with $\epsilon \ll 1$, there is no expansion of $h_{ik}$ into series.
Index ``$B$'' is introduced to denote background quantities in the Isaacson procedure after averaging.
 From the start he considers two scales, first, the wavelength of gravitational ripple $\lambda$. Second, the characteristic scale $L$ of
a spacetime, the curvature value of which is calculated with taking into account the back-reaction of the averaged energy-momentum.
Namely, such a spacetime is used as a background spacetime for perturbations in the Isaacson scheme.
Thus, vacuum Einstein equations $G_{ik} = 0$ after decomposition are derived by Isaacson  as
\begin{eqnarray}
G^B_{ik} + G^{(1)}_{ik} +G^{(2)}_{ik} = 0,
\label{GR.eq.note}
\end{eqnarray}
where $G^B_{ik}$ is the Einstein tensor for the averaged background with the metric $g^B_{ik}$. The second term in (\ref{GR.eq.note})  is linear in perturbations $h_{ik}$ on the averaged background, it defines the equation
\begin{eqnarray}
G^{(1)}_{ik}  = 0
\label{GR.G1.note}
\end{eqnarray}
that is just the gravitational wave equation determining $h_{ik}$. The third term in (\ref{GR.eq.note}) is quadratic in perturbations on the averaged background. Finally, averaging (\ref{GR.eq.note}) gives
\begin{eqnarray}
G^B_{ik} = - \langle G^{(2)}_{ik}\rangle = 8\pi G t^B_{ik},
\label{GR.eq+.note}
\end{eqnarray}
where $G$ is the gravitational constant, the units correspond to a choice $c=1$. The quantity $t^B_{ik}$ is just the averaged energy-momentum tensor of the gravitational waves in regions of the scale $S$ for which $S\gg\lambda$. Equations (\ref{GR.G1.note}) and (\ref{GR.eq+.note}) have to be solved simultaneously.

The equation (\ref{GR.eq+.note}) gives a possibility to connect parameters of the averaged background and of the gravitational wave as
\begin{eqnarray}
O\left(\frac{1}{L^2}\right) = O\left(\frac{\epsilon^2}{\lambda^2}\right).
\label{L.lambda}
\end{eqnarray}
Thus,  Isaacson's scheme leads to the relation $\lambda \ll L$ that clearly supports the notion ``high frequency''. Note that the scales $S$ and $L$ really are not connected from the start. For $S$ one has only $S\gg\lambda$, whereas $L$ is defined by (\ref{L.lambda}) connecting to the gravitational wave parameters. However,  the restriction $S\gg L$ is not possible because  in this case regions with the scale $L$ looks as a ripple with respect to scale $S$ that contradicts to the Isaacson averaging procedure.

Because the aforementioned scales are relative ones Isaacson uses a possibility to concretize a one of the them. He chooses $L=O(1)$, then by (\ref{L.lambda}) it turns out $\lambda = O(\epsilon)$.   As a result,
\begin{eqnarray}
g_{ik}^{B}\sim\partial_n g_{ik}^{B}\sim\partial_n\partial_m g_{ik}^{B} =O(1),~~h_{ik}=O(\epsilon),~~\partial_n h_{ik}=O(1),~~\partial_{m}\partial_{n} h_{ik}=O(\epsilon^{-1})
\label{a.3}
\end{eqnarray}
that leads to
 $G^B_{ik} = O(1)$ with $t^B_{ik} = O(1)$, and  $G^{(1)}_{ik} = O(\epsilon^{-1})$.

\subsection{Isaacson's procedure in $f(R)$ theory on dS background}

In \cite{TP1}, we have generalized the Isaacson procedure to be applied in the $f(R)$ theory with the use of the Brill-Hartle average method \cite{BH1}.
The action for $f(R)$ theory in vacuum \cite{FT1} is
\begin{equation}
S=\frac{1}{16\pi G}\int d^4x \sqrt{-g}f(R),
\label{a.1}
\end{equation}
where $R$ is the curvature scalar constructed with the metric $g_{ik}$ in the usual way in a pseudo-Riemannian space-time; $f(R)$ is enough smooth scalar function.  The equations of motion (field equations) corresponding to (\ref{a.1}) are
\begin{equation}
J_{ik}\equiv f'R_{ik} -\frac{1}{2}fg_{ik} +g_{ik}\Box f' -\nabla_i\nabla_k f' =0,
\label{a.2}
\end{equation}
where prime means a differentiation with respect to argument. Here and below, the Riemannian tensor $R^m{}_{ink}$,  the Ricci tensor $R_{ik}$ and the covariant derivatives $\nabla_i$ are constructed with the use of $g_{ik}$; $\Box \equiv \nabla_i\nabla^i$.

Parameters of the model are systemized as follows. Gravitational perturbations again are defined as in (\ref{decomposition}), and  gravitational ripple is defined with the amplitude of the order $\epsilon \ll 1$ and the wavelength $\lambda$. Thus,
\begin{equation}
h_{ik} = O(\epsilon);\qquad\partial_n h_{ik} = O\left( \epsilon/\lambda\right);\qquad \partial_m\partial_n  h_{ik}= O\left( \epsilon/\lambda^2\right); \ldots .
\label{h.lambda}
\end{equation}

To obtain the effective energy-momentum of gravitational waves we consider expansions of the quantity $J_{ik}$  in (\ref{a.2}).
Analogously to (\ref{GR.eq.note}) we expand the equation (\ref{a.2}) on the averaged background spacetime (created with taking into account a back-reaction) with the metric $g^{B}_{ik}$:
\begin{eqnarray}
J_{ik} &=& J^{B}_{ik} +  J^{(1)}_{ik} + J^{(2)}_{ik} = 0.
\label{J.B.1.2}
\end{eqnarray}
The expression $J^{(1)}_{ik}$  is linear in perturbations $h_{ik}$ on an arbitrary background. Analogously to (\ref{GR.G1.note}) we have the linear equation
\begin{eqnarray}
J^{(1)}_{ik}  = 0.
\label{f(R).J1}
\end{eqnarray}
Averaging (\ref{J.B.1.2}) on the scale $S \gg \lambda$ gives the analog of (\ref{GR.eq+.note}):
\begin{eqnarray}
 J^{B}_{ik} =-\langle J^{(2)}_{ik} \rangle & =& 8\pi G t^{B}_{ik}.
\label{J.1.2.angle}
\end{eqnarray}
Tensor $t^B_{ik}$ is the averaged energy-momentum tensor of the gravitational waves of high frequency, and $g^{B}_{ik}$ satisfies the non-vacuum equation (\ref{J.1.2.angle}). Besides, equations (\ref{f(R).J1}) and (\ref{J.1.2.angle}) have to be solved simultaneously.

However, our scheme in \cite{TP1} differs from the original Isaacson's scheme \cite{A1,A2}. The main point is that we introduce the initial background spacetime. It is the dS space with characteristic length $\ell$ that is the radius of the cosmological horizon. The quantities related to dS space are marked by ``$(0)$''. The dS space satisfies the equation (\ref{a.2}) for the background
\begin{equation}
J_{ik}^{(0)} = 0.
\label{J_0}
\end{equation}
The characteristic length $\ell$ determines the next relations
\begin{eqnarray}
g_{ik}^{(0)}=O(1),~~~\partial_ng_{ik}^{(0)} =O(\ell^{-1}),~~~\partial_{m}\partial_{n} g_{ik}^{(0)}=O(\ell^{-2}).
\label{b.5}
\end{eqnarray}
 \begin{eqnarray}
 R^{(0)}_{imkn} &=& \frac{1}{\ell^2}\Big(  g^{(0)}_{ik}g^{(0)}_{mn}- g^{(0)}_{mk}g^{(0)}_{in}\Big) = \frac{1}{6}\frac{f_0}{f'_0}\Big(  g^{(0)}_{ik}g^{(0)}_{mn}- g^{(0)}_{mk}g^{(0)}_{in}\Big),\nonumber\\
  R^{(0)}_{ik} &=&  \frac{3}{\ell^2}g^{(0)}_{ik} \frac{1}{2}\frac{f_0}{f'_0}g^{(0)}_{ik},\nonumber\\
   R^{(0)} &=& \frac{12}{\ell^2} = 2\frac{f_0}{f'_0}.
\label{R0_ell}
\end{eqnarray}
where
$f_0 = f(R^{(0)})$, $f'_0 = f'(R^{(0)})$, etc., are constant quantities. It is evidently that $S$ cannot exceed the radius of cosmological horizon, thus  $S := \lambda\ll S \leq \ell$ that leads to $\lambda\ll  \ell$.
In \cite{TP1}, the averaging procedure was applied just on the dS background $g^{(0)}_{ik}$ that significantly simpler than the one on the averaged background  $g^{B}_{ik}$.

Of course, it was necessary to compare the averaging results for both the cases. Following this requirement we have introduced the ``shift'':
\begin{eqnarray}
\tilde g^{B}_{ik}=g^{B}_{ik}- g^{(0)}_{ik}
\label{hB.phiB}
\end{eqnarray}
with its own characteristic scale ${\cal L}$:
\begin{eqnarray}
\tilde g_{ik}^{(0)}=O(1),~~~\partial_n\tilde g_{ik}^{(0)} =O({\cal L}^{-1}),~~~\partial_{m}\partial_{n} \tilde g_{ik}^{(0)}=O({\cal L}^{-2}).
\label{b.5.B}
\end{eqnarray}
Because $g^{B}_{ik}=\tilde g^{B}_{ik}+ g^{(0)}_{ik}$ one states from (\ref{J.1.2.angle}) the relation between the parameters of the model that generalized
(\ref{L.lambda}):
\begin{eqnarray}
O\left(\frac{ 1}{\ell{\cal L}}\right)+O\left(\frac{1}{{\cal L}^2}\right) = O\left(\frac{\epsilon^2}{\lambda^2}\right).
\label{L.lambda+}
\end{eqnarray}
The order $O({1}/{{\ell}^2})$ is not included because it is devoted to $J^{(0)}_{ik}$ satisfying (\ref{J_0}).

For the case ${\cal L}\gg \ell$ the equation (\ref{L.lambda+}) gives the connection of the parameters as $\epsilon^2\ell{\cal L} \sim \lambda^2$, whereas  ${\cal L}\leq \ell$ leads to $\epsilon{\cal L} \sim \lambda$. By the aforementioned reasons $S\leq {\cal L}$, thus $\lambda \ll {\cal L}$.
 Because the scales $\lambda$ and ${\cal L}$ are relative ones we have a possibility to choose one of them as a unit. Unlike \cite{A1,A2},
 was choose $\lambda \sim O(1)$. Then, (\ref{h.lambda}) transforms to
\begin{equation}
h_{ik} = O(\epsilon);\qquad\partial_n h_{ik} = O\left( \epsilon\right);\qquad \partial_m\partial_n  h_{ik}= O\left( \epsilon\right); \ldots .
\label{h.lambda+}
\end{equation}
Such an assumption significantly simplifies calculations comparisons. For example,
  we have important inequalities $\ell \gg 1$  and ${\cal L}\gg 1$. By the above conditions, we obtain in \cite{TP1} that the energy-momentum tensors averaged on the dS background $t^{(dS)}_{ik}$ and $t^{B}_{ik}$ in (\ref{J.1.2.angle}) coincide in the lower order $\epsilon^2$:
\begin{equation}
t^{(dS)}_{ik}=t_{ik}^{B} = \frac{f'_0}{32\pi G}
\left(\partial_ih^{nm}\partial_kh_{nm} +\frac{1}{8}  \partial_i h  \partial_k h  \right) = O(\epsilon^{2}).
\label{a.33B}
\end{equation}
However, obtaining $t^{(dS)}_{ik}$ is significantly simpler. Note, here, that because $f(R)$ theory is equivalent to a scalar-tensor theory the trace $h$ cannot be removed by gauge transformations.

\subsection{Second order metric perturbations and a possible contribution into the effective energy-momentum}

%%%%%%%%%%%%%%%%%%%%%%%%%

To obtain (\ref{a.33B}) in \cite{TP1} we did not considered the question how the second order metric perturbations can influence on the main order in (\ref{a.33B}). To provide such a consideration one has to use the decomposition
\begin{equation}
g_{ik}=g_{ik}^{(0)}+ h_{ik}+u_{ik},
\label{b.1}
\end{equation}
where we imply $g_{ik}^{(0)}=O(1)$, $h_{ik}=O(\epsilon)$ and $u_{ik}=O(\epsilon^2)$. Our main goal is to estimate a contribution of $u_{ik}$ into (\ref{a.33B}).

Thus, here, we develop formalism developed in \cite{TP1} and described shortly in previous subsection with the same restrictions and conditions. This means that the averaging procedure is provided on the dS background satisfying (\ref{b.5}) and (\ref{R0_ell}).
We also imply that $u_{ik}$ has its own scale $\lambda_1$, which is has the similar order with the main part of perturbation $\lambda_1\sim\lambda = O(1)$.  As a result, following (\ref{h.lambda+}) we have
\begin{eqnarray}
h_{ik} &\sim & \partial_n h_{ik} \sim \partial_{m}\partial_{n} h_{ik}\sim \ldots \sim \epsilon,
\label{b.2}\\
u_{ik} &\sim & \partial_n u_{ik} \sim \partial_{m}\partial_{n} u_{ik}\sim \ldots \sim \epsilon^2.
\label{b.3}
\end{eqnarray}
Besides, we keep in mind $\ell \gg 1$. Further using the evident relation $ g^{il} g_{lk}= \delta^i_k$, we derive the expansion of the contravariant metric components (rise and down of indexes are produced by using background metric):
\begin{equation}
g^{ik}= g^{ik\,(0)}- h^{ik} + h^{il}h_l^k -u^{ik} + O(\epsilon^3).
\label{b.4}
\end{equation}

\section{Second order perturbation in $f(R)$-gravity on de Sitter background}
\label{sect.3}

\subsection{Calculation of second order decomposition for Ricci tensor and Ricci scalar}

To achieve our goal we have to analyze $J^{(2)}_{ik}$ in (\ref{J.B.1.2}) in the framework of the decomposition (\ref{b.1}).
It needs to note that, from the one hand, we are forced to calculate all quantities from the very beginning, from another hand, the main expression (\ref{a.33B}) from our previous paper \cite{TP1} quadratic in derivatives of $h_{ik}$ is correct and we only need to supplement it by contribution from $u_{ik}$. Now, we derive necessary expressions of the second order. Thus, the second order connection reads as
\begin{equation}
\Gamma^{i\,(2)}_{kl}= -\frac{1}{2}h^{im}\left(\nabla_l^{(0)} h_{mk} + \nabla_k^{(0)} h_{ml} -\nabla_m^{(0)} h_{kl}\right) +\frac{1}{2}g^{im(0)}\left(\nabla_l^{(0)} u_{mk} + \nabla_k^{(0)} u_{ml} -\nabla_m^{(0)} u_{kl}\right),
\label{a.5}
\end{equation}
whereas expression for second order of Riemannian tensor formally is unchanged
\begin{equation}
R^{i\,(2)}_{~klm}=\nabla_l^{(0)}  \Gamma^{i\,(2)}_{km}- \nabla_m^{(0)}  \Gamma^{i\,(2)}_{kl}   + \Gamma^{i\,(1)}_{nl}\Gamma^{n\,(1)}_{km} -\Gamma^{i\,(1)}_{nm}\Gamma^{n\,(1)}_{kl}.
\label{a.6}
\end{equation}
Since all interesting us terms are linear in $u_{kl}$ it is enough for our task to calculate the parts $ R_{ik}^{(2)}[u_{ik}]$ and $R^{(2)}[u_{ik}]$. It is easy to see that for Riemannian tensor we have
\begin{eqnarray}
R^{i\,(2)}_{~klm}[u_{ik}]=&&\frac{1}{2}g^{ip(0)}\left(\nabla_l^{(0)}\nabla_m^{(0)} u_{pk} + \nabla_l^{(0)}\nabla_k^{(0)} u_{pm} -\nabla_l^{(0)}\nabla_p^{(0)} u_{km}\right) \nonumber\\ &&-\frac{1}{2}g^{ip(0)}\left(\nabla_m^{(0)}\nabla_l^{(0)} u_{pk} + \nabla_m^{(0)}\nabla_k^{(0)} u_{pl} -\nabla_m^{(0)}\nabla_p^{(0)} u_{kl}\right),
\label{a.7}
\end{eqnarray}
and simplifying first and third indexes we find easily

\begin{eqnarray}
R^{(2)}_{km}[u_{ik}]=\frac{1}{2}\left(\nabla_l^{(0)}\nabla_m^{(0)} u^l_{k} + \nabla_l^{(0)}\nabla_k^{(0)} u_{m}^l -\Box^{(0)} u_{km} - \nabla_m^{(0)}\nabla_k^{(0)} u \right),
\label{a.8}
\end{eqnarray}
where $u$ is the trace $u\equiv g^{ik(0)}u_{ik}$. And finally contracting total decomposition of Ricci tensor with total metric we find
\begin{eqnarray}
R^{(2)}[u_{ik}]=\nabla_l^{(0)}\nabla_m^{(0)} u^{lm}  -\Box^{(0)} u -  u^{km} R^{(0)}_{km}.
\label{a.9}
\end{eqnarray}

\subsection{Averaging procedure}

Let us derive necessary here properties of the Brill-Hartle averaging scheme \cite{BH1}. First, averaging total derivative term produce zero:
$$
\big\langle  \partial_j A_{iklm} \big\rangle =0,
$$
which allows us to switch derivatives from one object to another under the averaging
$$
\Big\langle \partial_j \big ( A_{ik} B_{lm} \big ) \Big\rangle = 0 =  \Big\langle A_{ik} \partial_j   B_{lm} \Big\rangle  +\Big\langle B_{lm} \partial_j  A_{ik}  \Big\rangle.
$$
Second, the background values are not averaged, so
$$
\big\langle g^{(0)}_{ik} h_{lm} \big\rangle = g^{(0)}_{ik} \big\langle  h_{lm} \big\rangle.
$$
Third, it is important to note that averaging of (\ref{a.8}) and (\ref{a.9}) produce zero in leading $\epsilon^2$ order, and non-trivial terms will produce
$$
\Big\langle R^{(2)}_{km}[u_{ik}] \Big\rangle = O(\ell^{-1}\epsilon^2) +O(\ell^{-2}\epsilon^2),
$$
$$
\Big\langle R^{(2)}[u_{ik}] \Big\rangle = O(\ell^{-1}\epsilon^2) +O(\ell^{-2}\epsilon^2).
$$

\subsection{A structure of the second order term $J^{(2)}_{ik}$ in the vacuum equation in $f(R)$-gravity}

Decomposition of equation (\ref{a.2}) without taking into account $u_{ik}$ was described in our previous paper \cite{TP1} in detail. Of course, introducing of $u_{ik}$ can change only second order in (\ref{J.B.1.2}), which in general case reads
\begin{eqnarray}
J^{(2)}_{ik}= &&f'^{(0)}R_{ik}^{(2)} +f'^{(1)}R_{ik}^{(1)} +f'^{(2)}R_{ik}^{(0)} -\frac{1}{2}f^{(0)}u_{ik} -\frac{1}{2}f^{(1)}h_{ik}-\frac{1}{2}f^{(2)}g^{(0)}_{ik} +  \nonumber\\
 \nonumber\\
&& g^{(0)}_{ik}\Big(\Box f' \Big)^{(2)} + h_{ik}\Big(\Box f' \Big)^{(1)} + u_{ik}\Big(\Box f' \Big)^{(0)}-\Big(\nabla_i\nabla_k f'\Big)^{(2)},
\label{a.3}
\end{eqnarray}
where we use natural decomposition of function $f$ and its derivative $f'$
$$
f^{(1)}=f'_0R^{(1)},\,\,\,f^{(2)}=f'_0R^{(2)}+\frac{1}{2}f''_0(R^{(1)})^2,\,\,\,f'^{(1)}=f''_0R^{(1)},\,\,\,f'^{(2)}=f''_0R^{(2)}+\frac{1}{2}f'''_0(R^{(1)})^2.
$$

Remember that we are interested in calculation of possible contribution to second order of decomposition (\ref{a.3}) from second order part decomposition of metric $u_{ik}$. Thus we can ignore the second, fifth and eighth terms in (\ref{a.3}) because they are superposition of the first order which is unchange and can't produce something new.

Now let us discuss possible functions $f(R)$.
In the normal case function $f$ is implied by the order $O(\ell^{-2})$. Indeed for the most number of realistic functions decomposition looks like $f(R)=R+O(R^2)+..$ -- it is need to have GR as a low energy limit.

For instance, for Starobinsky model \cite{Star} we have

$$
f_S=R+\lambda R_s\bigg [\Big( 1+\frac{R^2}{R_s^2}\Big)^{-n} -1  \bigg ],~~~f_S=R-\frac{\lambda n}{R_s}R^2+\frac{1}{2}\frac{\lambda n(n+1)}{R_s^3}R^4 -\frac{1}{6}\frac{\lambda n(n^2+3n+2)}{R_s^5}R^6+...,
$$
and since on de Sitter background we have $R_0=12/\ell^2$ in the leading order we have $f_0=O(\ell^{-2})$, whereas for its derivative we find $f'_0=O(1)$ and $f''_0=O(1)$.

\subsection{Contribution to energy-momentum tensor from second order metric perturbation}

Here, keeping in mind all the aforementioned restrictions, conditions and properties under averaging on the dS background (among them $\ell^{-1}\ll 1$), we average all the items in (\ref{a.3}).
Remind also that leading order in energy-momentum tensor (\ref{a.33B})  is $\epsilon^2$, so we can neglect by the all terms weaker it. We repeat that we do not consider second, fifth and eighth terms because they are quadratic in $h_{ik}$  and their derivatives, and, thus, already have been taken into account in (\ref{a.33B}). We continue with the rest terms.

The first term produces
$$
\langle f'^{(0)}R_{ik}^{(2)} \rangle = O(\ell^{-1}\epsilon^2) +O(\ell^{-2}\epsilon^2);
$$
the third term produces
$$
\langle f'^{(2)}[u_{ik}]R_{ik}^{(0)} \rangle =  \langle f''^{(0)}R^{(2)}[u_{ik}]R_{ik}^{(0)} \rangle= O(\ell^{-3}\epsilon^2) +O(\ell^{-4}\epsilon^2);
$$
the forth term produces
$$
\langle f^{(0)}u_{ik}  \rangle = O(\ell^{-2}\epsilon^2);
$$
the sixth term produce
$$
\langle f^{(2)}[u_{ik}]g^{(0)}_{ik} \rangle =  \langle f'^{(0)}R^{(2)}[u_{ik}]g^{(0)}_{ik} \rangle=O(\ell^{-1}\epsilon^2) +O(\ell^{-2}\epsilon^2);
$$
it is evidently, that all these terms may be neglected.

Concerning the seventh term, decomposing total expression
$$\Box f'\equiv g^{ik}\nabla_i\nabla_kf'=g^{ik}\partial_i\partial_k f' - g^{ik}\Gamma^p_{ik} \partial_p f'
$$
we obtain the next expression
\begin{eqnarray*}
\Big(\Box f' \Big)^{(2)}=&&g^{ik(0)}\partial_i\partial_k f'^{(2)} -h^{ik}\partial_i\partial_k f'^{(1)} - g^{ik(0)}\Gamma^{p(0)}_{ik}\partial_pf'^{(2)} -g^{ik(0)}\Gamma^{p(1)}_{ik}\partial_pf'^{(1)}+ h^{ik}\Gamma^{p(0)}_{ik}\partial_pf'^{(1)}  -u^{ik}\partial_i\partial_k f'^{(0)} \nonumber\\
&&+ h^{il}h^k_l\partial_i\partial_k f'^{(0)}-g^{ik(0)}\Gamma^{p(2)}_{ik}\partial_pf'^{(0)} +h^{ik}\Gamma^{p(1)}_{ik}\partial_pf'^{(0)} - h^{il}h^k_l\Gamma^{p(0)}_{ik}\partial_pf'^{(0)} +u^{ik}\Gamma^{p(0)}_{ik}\partial_pf'^{(0)} .
\label{a.4}
\end{eqnarray*}
We can see that last six terms here are equal to zero due to background relation, the average of the first and third terms transforms to averaging total derivatives leading to zero, and the remaining three terms do not produce any new contribution to final result following from $u_{ik}$.

Due to our background convention, the ninth term produces zero
$$
\langle u_{ik}\Big(\Box f' \Big)^{(0)} \rangle = 0.
$$

Concerning the tenth term,
expanding in obvious expression $\nabla_i\nabla_kf'=\partial_i\partial_k f' - \Gamma^p_{ik} \partial_p f'$ every term up to the second order we find easily
\begin{eqnarray*}
\Big(\nabla_i\nabla_k f'\Big)^{(2)}=\partial_i\partial_k f'^{(2)} -\Gamma^{p(1)}_{ik}\partial_pf'^{(1)}-\Gamma^{p(2)}_{ik}\partial_pf'^{(0)} - \Gamma^{p(0)}_{ik}\partial_pf'^{(2)}.
\label{a.4}
\end{eqnarray*}
Here,  the first and last terms do not produce contribution to the final answer because their average leads to averaging total derivatives giving zero, the third term is equal to zero due to the background convention and the second one does not produce any new contribution.

Finally we conclude that there is no contribution to this term from $u_{ik}$.

Our result depends crucially on the assumption $\lambda_1 \sim \lambda =O(1)$, where $\lambda_1$ is wavelength of the secondary metric perturbation. The assumption $\lambda_1 > \lambda =O(1)$ only strengthens our result, because in this case instead of (\ref{b.3}) we have
$$
u_{ik} \sim \epsilon^2;\qquad  \partial_n u_{ik} \sim \epsilon^2/\lambda_1;\qquad \partial_{m}\partial_{n} u_{ik}  \sim \epsilon^2/\lambda^2_1; \qquad\ldots .
$$
Our result can be really changed if we suppose  $\lambda_1 \leq    \ell^{-1}\ll 1$. However, this assumption means that the second order perturbation has a more high frequency than the main frequency. By this, the structure of the equation (\ref{f(R).J1}) can be changed that is not in the scope of the Isaacson procedure. Thus, we do not consider such a possibility.

\section{Conclusion}
\label{sect.4}

We have analyzed all possible contributions from the second order metric decomposition term $u_{ik}$ to the energy-momentum tensor on the dS background (\ref{a.33B}). We have found that all these contributions either are the total derivatives or have lower order then the leading terms in (\ref{a.33B}). By the conditions and restrictions in \cite{TP1}, the same conclusion is related to the average procedure on the averaged background $g^B_{ik}$. Thus we can neglect by the second-order term in decomposition of metric tensor without loss of generality. It is important to remark that this assertion takes a place for an arbitrary characteristic length $\lambda_1$ of the ripple $u_{ik}$, which does not destruct the Isaacson procedure. Finally we remark that the results have been obtained in the relatively simple way and form due to the choice of the scales with $\lambda = O(1)$. In the case of Isaacson's choice, when one has to set ${\cal L} = O(1)$ instead of  $\lambda = O(1)$, the same results could be obtained in a significantly more cumbersome way and form.

\bigskip

{\bf Acknowledgments.} The study of AP was conducted under the state assignment of Lomonosov Moscow State University.

\end{document}